\documentclass[reprint,aps,pra,superscriptaddress,floatfix,onecolumn,notitlepage]{revtex4-1}
\pdfoutput=1
\usepackage[utf8]{inputenc}
\usepackage[T1]{fontenc}

\usepackage{xparse}

\usepackage{acronym}
\usepackage{bbold}
\usepackage{bm}
\usepackage{braket}
\usepackage{mathtools}
\usepackage{amsfonts}
\usepackage[shortlabels]{enumitem}

\usepackage{array}
\newcolumntype{C}{>{\(}c<{\)}}

\NewDocumentCommand\Cfi{}{\mathcal{I}}
\NewDocumentCommand\cfi{om}{\Cfi(#2)}
\NewDocumentCommand\Qfi{}{\mathcal{J}}
\NewDocumentCommand\qfi{om}{\Qfi\IfValueT{#1}{_{#1}}(#2)}

\newcommand\ketbra[1]{\ket{#1}\bra{#1}}

\newcommand\expect[1]{\langle #1 \rangle}
\newcommand\op[1]{\hat{#1}}

\newcommand\psiNOON{\psi_{\mathrm{N00N}}}
\newcommand\psiONs{\psi_{\mathrm{0N}}^{\otimes 2}}
\newcommand\psiON{\psi_{\mathrm{0N}}}
\newcommand\psiONN{\psi_{\mathrm{0NN}}}
\newcommand\rhoONs{\rho_{\psiONs}}

\newcommand\rhoONN{\rho_{\psiONN}}

\newcommand\nTot{\op{n}_{\mathrm{Tot}}}
\newcommand\onN{N}
\newcommand\avgN{\expect{N}}
\newcommand\avgSqN{\expect{N^2}}

\DeclarePairedDelimiter\Abs{\lvert}{\rvert}                                                                       
\newcommand\abs[1]{\Abs*{#1}}

\DeclareMathOperator{\Tr}{Tr}

\acrodef{CFI}{classical Fisher information}
\acrodef{CRB}{Cramér-Rao bound}
\acrodef{QCRB}{quantum Cramér-Rao bound}
\acrodef{QFI}{quantum Fisher information}
\acrodef{SNL}{shot-noise limit}
\acrodef{SLD}{symmetric logarithmic derivative}

\usepackage{hyperref}

\begin{document}
\title{Average number is an insufficient metric for interferometry}

\author{Dominic Branford}
\affiliation{Institute of Photonics and Quantum Sciences, Heriot-Watt University, Edinburgh, EH14 4AS, United Kingdom}
\author{Jesús Rubio}
\affiliation{Department of Physics and Astronomy, University of Exeter, Stocker Road, Exeter, EX4 4QL, United Kingdom}

\date{\today}

\begin{abstract}
We argue that analysing schemes for metrology solely in terms of the average particle number can obscure the number of particles effectively used in informative events.
For a number of states we demonstrate that, in both frequentist and Bayesian frameworks, the average number of a state can essentially be decoupled from the aspects of the total number distribution associated with any metrological advantage.
\end{abstract}

\maketitle

\section{Introduction}

Quantum sensing techniques offer the possibility of improved precision per (average) particle~\cite{giovannetti_quantum_2006}.
In phase estimation, the precision \( \Delta \varphi \gtrsim 1/\sqrt{\avgN} \)---the \ac{SNL}---available from repeated use of single photons or classical light can be surpassed, instead attaining \( \Delta\varphi \gtrsim 1/\avgN \)---Heisenberg scaling---through quantum correlations such as entanglement and squeezing~\cite{giovannetti_quantum_2006,lang_optimal_2014,toth_quantum_2014,demkowicz-dobrzanski_quantum_2015,sahota_quantum_2015}.

Only being a quadratic improvement, the tens of photons found in the quantum state-of-art~\cite{vahlbruch_detection_2016,wang_boson_2019} cannot compare directly with the intensity of classical laser light.
Indeed, only recently has the shot-noise limit been surpassed in absolute terms~\cite{slussarenko_unconditional_2017}.
While latest generation gravitational wave detectors utilise squeezed light, the current frequency-independent squeezing does not decrease quantum noise in absolute terms%
~\footnote{Radiation-pressure effects in these detectors introduce a trade-off where a maximum sensitivity is reached beyond which the negative contribution from back-action outweighs the reduction in shot-noise without frequency-dependent measurements and/or squeezing~\cite{kimble_conversion_2001}},
but can beneﬁt certain gravitational wave frequencies, or reduce the required classical light and associated classical noise contributions.

Instead, the argument for more immediate applications of the quantum metrology toolbox have focused on the idea of delicate samples~\cite{wolfgramm_entanglement-enhanced_2013,taylor_biological_2013,taylor_quantum_2016,triginer_garces_quantum-enhanced_2020,casacio_quantum-enhanced_2021,jolly2021}.
While a piece of glass or a waveplate may be robust to strong laser light, biological systems may be expected to suffer photodamage~\cite{cole_live-cell_2014}.
Such biological limits give cause to want to constrain the amount of (damaging) light passing through the sample itself.

These constraints can entail limiting the total, average, or single-shot number of photons the sample is exposed to.
While optimal states in a truncated Fock-space setting are known~\cite{demkowicz-dobrzanski_optimal_2011,lang_optimal_2014}, this strict setting necessitates excluding otherwise reasonable states---that can likely be generated in a more scalable fashion---including squeezed vacuum which have small but non-zero overlap with large Fock states.
As a result, even when the delicate sample motivation is invoked, average number is often exclusively used as a resource.

For states without a fixed total number, the quadratic-in-\( \avgN \) variance of Heisenberg scaling is not an absolute limit and a greater per-shot precision for fixed \( \avgN \) has been predicted under local estimation~\cite{rivas_sub-heisenberg_2012,zhang_unbounded_2012}, although the regime where local estimation applies is not easily accessed in such settings~\cite{tsang_ziv-zakai_2012,giovannetti_sub-heisenberg_2012,berry_optimal_2012,pezze_sub-heisenberg_2013}.
A bound \( \Delta\varphi \gtrsim 1/\sqrt{\avgSqN} \) has been proposed as an alternate or complementary limit~\cite{hofmann_all_2009,hyllus_entanglement_2010,pezze_phase-sensitivity_2015} to the \( \Delta\varphi \gtrsim 1/\avgN \) Heisenberg scaling.

In this work we see that average number \( \avgN \) is not only insufficient to construct an absolute bound to phase estimation but that it can bear no relevance to either the metrological performance of variable number states or the number of particles that may be passing through a sample on any given shot.
This leads us to conclude that average number is not just an occasionally imperfect proxy for the cost of certain states, but a potentially dangerous misnomer as to the exposure a sample may receive.

\section{Interferometry}
\label{sec:interferometry}

\begin{figure}[htbp]
\centering
\includegraphics{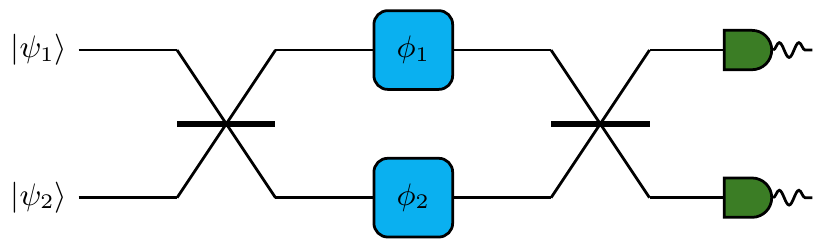}
\caption{Mach-Zehnder Interferometry scheme with two one-mode states incident directly on the local phases.
The initial beam splitter is sometimes omitted as in Figs.~\ref{fig:psi_noon} and~\ref{fig:psi_on}, particularly for two-mode input states.
Separable inputs are normally considered when the initial beam splitter is present.}
\label{fig:psi}
\end{figure}

Typically, analyses of interferometry, such as the full Mach-Zehnder scheme illustrated in Fig.~\ref{fig:psi}, focus on maximising the precision for a given average number of photons~\cite{lang_optimal_2014,sahota_quantum_2015}.
While there are families of fixed number states---particularly N00N~\cite{mitchell_super-resolving_2004,giovannetti_advances_2011} and Holland-Burnett~\cite{holland_interferometric_1993,lang_optimal_2014} states which always have exactly the same number of photons---in general variable number states are of interest~\cite{lang_optimal_2014,demkowicz-dobrzanski_quantum_2015}.
Among states without a fixed total number, Gaussian probes such as squeezed states can be considered to have relatively normal photon number distributions~\cite{lvovsky_squeezed_2015,demkowicz-dobrzanski_quantum_2015}.
However, states with more irregular photon number distributions have also been considered~\cite{rivas_sub-heisenberg_2012,zhang_unbounded_2012,lee_using_2019}.

In this section we introduce the framework of local estimation based around the \ac{CRB}~\cite{kay_fundamentals_1998} which we use in our first pass of the problem.
We revisit this under a Bayesian framework in Sec.~\ref{sec:bayes} to allow a more general control of the initial level of ignorance~\cite{jarzyna_true_2015}.

\subsection{Precision in quantum mechanics}

For a given probe state \( \rho \) and measurement \( \bm{\Pi} \) the precision attainable by any \emph{unbiased} estimate of a parameter is bounded by the \ac{CRB}.
Measurement-independent lower bounds on the \ac{CRB} then exist as \acp{QCRB}, which depend only on the quantum state.
For single-parameter estimation the \ac{SLD} \ac{QCRB} is the relevant \ac{QCRB}, as there exists a measurement which can satisfy the \ac{QCRB} inequality.
This forms a hierarchy of inequalities bounding the variance of the estimator~\cite{braunstein_statistical_1994,paris_quantum_2009}
\begin{equation}
	(\Delta \varphi)^2 \geq \frac{1}{\nu\cfi{\rho(\varphi);\bm{\Pi}}} \geq \frac{1}{\nu\qfi{\rho(\varphi)}},
    \label{eq:crbs}
\end{equation}
where \( \nu \) is the number of repetitions, \( \cfi{\rho(\varphi);\bm{\Pi}} \) the (state- and measurement-dependent) \ac{CFI}, and \( \qfi{\rho(\varphi)} \) the (state- but not measurement-dependent) \ac{SLD} \ac{QFI} (henceforth \ac{QFI}).

\subsection{Precision in interferometry}

For pure states, the \ac{QFI} for estimation of a phase shift \( { \varphi = \phi_1-\phi_2 } \) is straightforwardly given by~\cite{paris_quantum_2009,lang_optimal_2013,demkowicz-dobrzanski_quantum_2015}
\begin{equation}
	\qfi[\varphi]{\ket{\psi}} = \expect{(\op{n}_1-\op{n}_2)^2} - \expect{\op{n}_1-\op{n}_2}^2 
	\label{eq:phase_qfi}
\end{equation}
from the generating Hamiltonian \( K = (\op{n}_1-\op{n}_2) / 2 \).
The \ac{QCRB} then gives a bound
\begin{equation}
    (\Delta \varphi)^2 \geq \frac{1}{\nu \left[ \expect{(\op{n}_1-\op{n}_2)^2} - \expect{\op{n}_1-\op{n}_2}^2 \right]}.
    \label{eq:qcrb_generator}
\end{equation}
For mixed states the required expressions can be far more involved~\cite{paris_quantum_2009,demkowicz-dobrzanski_quantum_2015,genoni_non-orthogonal_2019} and the mixed state \acp{QFI} necessary for this work are derived in App.~\ref{app:mixed_qfi}.

\begin{figure}[htb]
\centering
\includegraphics{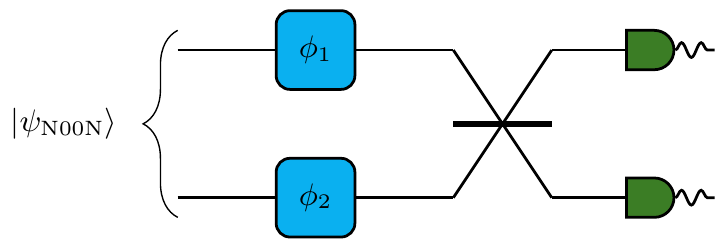}
\caption{Interferometry scheme with N00N state input.}
\label{fig:psi_noon}
\end{figure}

Among fixed-number states the maximum \ac{QFI} comes from the N00N state~\cite{giovannetti_quantum_2006,lang_optimal_2014}, 
which is a balanced superposition of all photons in one mode or the other
\begin{equation}
    \ket{\psiNOON} = \frac{\ket{N0}+\ket{0N}}{\sqrt{2}},
    \label{eq:noon}
\end{equation}
directly incident on the phase shifts as Fig.~\ref{fig:psi_noon}.
Using Eq.~\eqref{eq:phase_qfi}, \( \op{n}_1\ket{\psiNOON} = N \ket{N0} / \sqrt{2} \) and \( \op{n}_2\ket{\psiNOON} = N \ket{0N} / \sqrt{2} \) give
\begin{equation}
	\qfi{\ket{\psiNOON}} = N^2,
	\label{eq:qfi_noon}
\end{equation}
giving a Heisenberg-scaling precision \( (\Delta\varphi)^2 \geq 1/\nu N^2 \).

\subsection{Phase references}

The conventional measurement in optical interferometry is photon counting which is a phase-insensitive technique, although it can be used with linear optics to resolve relative phases it is insensitive to a local phase.
Where a state of indefinite photon number is used with phase-insensitive measurements the pure state (single-parameter) \ac{QCRB} is not necessarily sufficient to properly model the problem~\cite{jarzyna_quantum_2012,demkowicz-dobrzanski_quantum_2015}.

A single-mode state of indefinite number can---does, if pure---possess coherence between components of different total number which accumulate a local phase.
This phase is not directly accessible through phase-insensitive measurements---the probabilities \( \{ \abs{\braket{\psi|n}}^2 \} \) depend only on the magnitude of the amplitude in the Fock basis---without introducing (an) additional state(s) with a well-defined relative phase.
This is done both explicitly in the two-mode Mach-Zehnder or implicitly in the measurement as e.g.\ a local oscillator in homodyne and heterodyne detection~\cite{leonhardt_measuring_1995}.

While these phase-sensitive measurements are reasonable to consider in general, their availability fundamentally changes the formulation of the problem: it becomes capable to resolve a local phase shift with single-mode probe states~\cite{monras_optimal_2006,jarzyna_quantum_2012}.
As such a Mach-Zehnder scheme with phase-sensitive measurements may contain unnecessary overhead and perform worse than a phase-sensitive measurement on a single-mode probe state under an equivalent resource.

When we wish to explicitly consider phase-insensitive measurements (such as photon counting in Fig.~\ref{fig:psi}) the input probe state should be modified to erase coherences with respect to different degenerate eigenstate subspaces of the total number operators as~\cite{jarzyna_quantum_2012}
\begin{equation}
    \rho_{\psi} = \frac{1}{2\pi} \int\limits_{0}^{2\pi} \mathrm{d}\vartheta e^{i\vartheta(\op{n}_1+\op{n}_2)} \ketbra{\psi} e^{-i\vartheta(\op{n}_1+\op{n}_2)} ,
    \label{eq:phase_randomised}
\end{equation}
whence it follows \( e^{i\zeta (\op{n}_1+\op{n}_2)} \rho_{\psi} e^{-i\zeta (\op{n}_1+\op{n}_2)} = \rho_{\psi} \).
Unless \( \ket{\psi} \) is a state of fixed total number then \( \rho_{\psi} \) is necessarily a mixed state and \( \qfi{\rho_{\psi}} \) cannot be evaluated by Eq.~\eqref{eq:phase_qfi}, however this does not preclude \( \qfi{\rho_{\psi}} = \qfi{\ket{\psi}} \) which is in fact true for all the states we consider in the following section.

\section{Obfuscation through average number}
\label{sec:excess}

\begin{figure}[htb]
\centering
\includegraphics{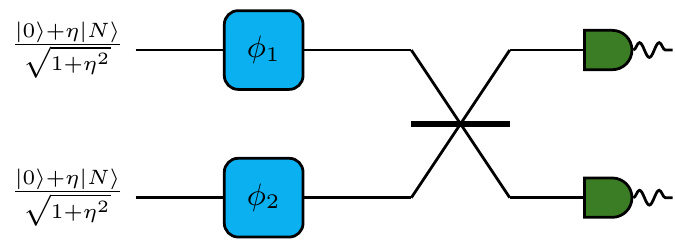}
\caption{Interferometry scheme with pair vacuum-Fock superpositions as input.}
\label{fig:psi_on}
\end{figure}

We first consider the vacuum-Fock superposition states
\begin{equation}
\ket{\psiONs} = \frac{(\ket{0}+\eta\ket{\onN})^{\otimes \mathrlap{2}}}{1+\eta^2}
= \frac{\ket{0}+\eta(\ket{\onN 0}+\ket{0 \onN}) + \eta^2\ket{\onN \onN}}{1+\eta^2}
\label{eq:psi_ons}
\end{equation}
states which are incident directly on a relative phase as in Fig.~\ref{fig:psi_on}.
These can be considered a Fock state analog of the Rivas-Luis state~\cite{rivas_sub-heisenberg_2012} to which we would expect the main ideas of this section to generalise.

These \( \ket{\psiON} \) states are a simple example to illustrate why we should be concerned with average number representing the entire description, ``danger'', and power of a variable number state.
We do not propose them as viable probe states: the generation of such a state \( \ket{\psiON} \) can be expected to be demanding and difficult to scale~\cite{loredo_generation_2019}, before worrying about the relative magnitudes of the amplitudes or the simultaneous generation of two copies with fixed and known relative phase.
Nor as worthwhile probe states: as well as numerous information-theoretic concerns made previously~\cite{tsang_ziv-zakai_2012,berry_optimal_2012},
we will demonstrate these in fact offer no advantage over an \emph{equivalent} N00N state.

\subsection{Arbitrary precision from fixed average number states}
\label{sec:arbitrary}

On average \( \ket{\psiONs} \) has
\begin{equation}
    \avgN = \braket{ \psiONs | \nTot | \psiONs } = \frac{ \eta^2 \onN }{ 2(1+\eta^2) }
    \label{eq:n_on}
\end{equation}
photons.
The \ac{QFI} for \( \varphi = \phi_1-\phi_2 \) can be evaluated from Eq.~\eqref{eq:phase_qfi} to
\begin{equation}
	\qfi[\varphi]{\ket{\psiONs}} = \frac{2\eta^2 \onN^2}{(1+\eta^2)^2}.
   \label{eq:qfi_on}
\end{equation}
Eq.~\eqref{eq:qfi_on} is quadratic in \( \onN \) while the average photon number is linear in \( \onN \) and so we attain Heisenberg scaling for a fixed \( \eta \).

However, rewriting Eq.~\eqref{eq:qfi_on} in terms of \( \onN \) and \( \expect{N} \)---by solving Eq.~\eqref{eq:n_on} for \( \eta \), namely \( \eta = \sqrt{ \frac{ \expect{N} }{ 2\onN - \expect{N} } } \) which is valid for \( \onN > \expect{N}/2 \)---we have
\begin{equation}
	\qfi[\varphi]{\ket{\psiONs}} =
   \expect{N} \left( \onN - \frac{\expect{N}}{2} \right),
   \label{eq:qfi_on_fixed_nbar}
\end{equation}
which can be made arbitrarily large for any fixed \( \expect{N} \) by making \( \onN \) arbitrarily large and adjusting \( \eta \) accordingly.

Observations of such large \acp{QFI} have been made before~\cite{rivas_sub-heisenberg_2012,zhang_unbounded_2012}. 
One problem that should be anticipated from similar states is in finite trials due to issues with the necessary repetitions~\cite{giovannetti_sub-heisenberg_2012,tsang_ziv-zakai_2012,luis_breaking_2017,rubio_quantum_2019} or necessary prior knowledge~\cite{hall_does_2012,hayashi_resolving_2018,rubio_bayesian_2020}.
While these limitations amply motivate alternative probe states, here we will show that the apparent surpassing of Heisenberg scaling in \( \avgN \) can be explained also within local estimation itself. 

With a N00N state the same precision can be reached with
\(
    \sqrt{2} \eta \onN / (1+\eta^2)
\)
photons (or at least approximated with the nearest integers).
For \( \eta \geq 1/\sqrt{2} \) this requires fewer photons on average, however when \( \eta \ll 1 \) (which is the case when Eq.~\eqref{eq:qfi_on_fixed_nbar} is made to substantially exceed Heisenberg scaling) the N00N state requires far more photons than the average number in the vacuum-Fock superpositions for the same precision.
This suggests the vacuum-Fock superposition (with \( \eta \ll 1 \)) should be preferred over a N00N state with equivalent \ac{QFI} due to the lower average number.
We will argue that, even if both states can be easily generated, such a predilection is ill-founded and may be deeply misguided.

\subsection{Excessive damage through detection events}
\label{sec:counting}

We can divide the potential measurement outcomes into three groups: a.\ neither detector clicks, b.\ both detectors click a total of \( N \) times, c.\ both detectors click a total of \( 2N \) times.
The latter two events subdivide into a combinatoric wealth of different distributions between the two detectors.
Already we can see that when we register photons as having passed through the sample, the number thereof can be entirely detached from \( \expect{N} \), given in Eq.~\eqref{eq:n_on}, which vanishes in the \( \eta \ll 1 \) limit even though \( N \) remains comparatively large.

In each case we can identify distinct amounts of information which will be returned:
\begin{enumerate}[a.,noitemsep]
\item no photons passed through the relative phase, no clicks, no information;
\item \( N \) photons passed through the relative phase shift in superposition and their resulting distribution between the two detectors is parameter-dependent;
\item \( N \) photons passed through each arm (\( 2N \) total) and so no relative phase was accumulated.
\end{enumerate}
In the first case one may be disappointed as we learn precisely nothing, though any sample is unscathed.
In the second, \( N \) photons pass through the sample and the \ac{QFI} and their counts display parameter-dependence.
In the last though, \( 2N \) photons passed through the sample, yet the counting statistics are entirely uninformative.

Where photon numbers in the single-shot are of utmost concern the existence of such an event could destroy the sample or necessitate the experiment be run with lower average numbers at the expense of precision.
Yet this is not apparent if we rely on the average number to be a sufficient approximation for the number of photons in the single-shot.
Nor were we expecting to only gain information when at least \( N \)---rather than \( \avgN \)---photons pass through the sample.

\subsection{The role of coherences across total number}

We can consider the projection into the subspaces spanned by \( \{ \ket{00} \} \), \( \{ \ket{\onN 0} , \ket{0 \onN} \} \), and \( \{ \ket{NN} \} \)---the effective state to phase-insenstive measurements---of \( \ket{\psiONs} \) according to Eq.~\eqref{eq:phase_randomised}, namely \( \rho_{\psiONs} \).
This gives us the state
\begin{equation}
    \rhoONs = \frac{
    \ketbra{00} + 2\eta^2 \ketbra{\psiNOON}  + \eta^4 \ketbra{\onN\onN}
    }{(1+\eta^2)^2} .
    \label{eq:rho_preselect}
\end{equation}
The \ac{QFI} of \( \rhoONs \), evaluated in App.~\ref{app:mixed_qfi:noonlike} using the \ac{SLD}, is
\begin{equation}
    \qfi{\rhoONs} = \frac{2\eta^2N^2}{(1+\eta^2)^2},
    \label{eq:qfi_rhoONs}
\end{equation}
matching Eq.~\eqref{eq:qfi_on},
whence we see that the optimal measurement for the probe state \( \ket{\psiONs} \) is phase-insensitive.
Moreover, as only the \( \ket{\psiNOON} \) component in both \( \ket{\psiONs} \) and \( \rho_{\psiONs} \) acquires a \( \varphi \) dependence (\( e^{i\varphi(\op{n}_1-\op{n}_2)/2} \ket{MM} = \ket{MM} \)) we can identify that a sufficient measurement to attain the \ac{QCRB} is one which both attains the \ac{QCRB} for \( \ket{\psiNOON} \) and distinguishes between components of different total photon number---the photon counting illustrated in Fig.~\ref{fig:psi_on} is one such measurement.

This probabilistic mixture of the definite total-number components of \( \ket{\psiONs} \) is no longer mode-separable.
Now, the only coherence is found in the (mode-entangled) N00N state term, which was implicitly present in the expansion of the tensor product in Eq.~\eqref{eq:psi_ons}.
The appearance of this N00N state component is not coincidental but is responsible---alongside it's vanishing weighting for \( \eta \ll 1 \)---for the arbitrarily large \ac{QFI} of the \( \ket{\psiONs} \) state.

\subsection{A probabilistic alternative (interpretation)}

While we introduced \( \rhoONs \) in the previous section as the effective state seen by phase-insensitive measurements it is itself a valid probe state---one which is just as capable as \( \ket{\psiONs} \).
Indeed whether created independently or by a suitable collapse on \( \ket{\psiONs} \) it is informative to understanding the underpinnings of the better-than-Heisenberg behaviour displayed by both \( \ket{\psiONs} \) and \( \rhoONs \).

Convexity of the \ac{QFI} upperbounds the \ac{QFI} of any mixed state as~\cite{toth_extremal_2013,alipour_extended_2015,ng_spectrum_2016}
\begin{equation}
    \qfi{\sum\limits_k p_k \ketbra{\psi_k}} \leq \cfi{\{p_k\}} + \sum\limits_{k} p_k \qfi{\ketbra{\psi_k}},
    \label{eq:qfi_convexity}
\end{equation}
where the two terms are the \ac{CFI} of the probabilities attached to the pure state decomposition and the average of the \acp{QFI} of the associated pure states.

For \( \rhoONs \) using the decomposition given in Eq.~\eqref{eq:rho_preselect}, the convexity inequality is satisfied with \( \qfi{\psiNOON} = \onN^2 \) and \( p_{\psiNOON} = 2\eta^2/(1+\eta^2)^2 \), and \( \qfi{\ket{00}} = \qfi{\ket{NN}} = \cfi{\{ \frac{1}{(1+\eta^2)^2} , \frac{2\eta^2}{(1+\eta^2)^2} , \frac{\eta^4}{(1+\eta^2)^2} \}} = 0 \).
Thus using \( \rhoONs \) (\(\ket{\psiONs}\)) as probe state is metrologically equivalent to probabilistically using the equivalent N00N state with the same \( \onN \) and equivalent probability \( p_{\psiNOON} = \braket{ \psiNOON | \rhoONs | \psiNOON } = \abs{\braket{ \psiNOON | \psiONs }}^2 \).

This is not to say that \( \rhoONs \) and \( \psiONs \) are entirely equivalent.
If we prepare \( \rhoONs \) by probabilistically generating \( \ket{00} \), \( \ket{\psiNOON} \), or \( \ket{NN} \) or measuring total number of \( \psiONs \) we can technically improve on the strategy by not preparing or discarding the \( \ket{NN} \) component.
Without material change to the interferometry scheme using instead
\begin{equation}
    \rhoONN = \frac{1+\eta^4}{(1+\eta^2)^2} \ketbra{00} + \frac{2\eta^2}{(1+\eta^2)^2} \ketbra{\psiNOON},
    \label{eq:rho_probabilistc_on}
\end{equation}
has a negligible effect on \( \avgN \) with \( 2\eta^2 N/(1+\eta^2)^2 \) dominating \( 2\eta^4 N/(1+\eta^2)^2 \) when \( \eta \ll 1 \), but still has the same \ac{QFI} \( 2\eta^2 N/(1+\eta^2)^2 \).
More importantly though it disposes of the potentially damaging but metrologically useless \( 2N \) events, making a significant difference to the worst-case in spite of a vanishing change in \( \avgN \).

As \( \ket{\psiONs} \) is to \( \rhoONs \), a coherent form of \( \rhoONN \) can also be defined as
\begin{equation}
	\ket{\psiONN} = \frac{\sqrt{1+\eta^4}}{1+\eta^2} \ket{0} + \frac{\sqrt{2}\eta}{1+\eta^2} \frac{\ket{N0}+\ket{0N}}{\sqrt{2}},
    \label{eq:on_trunc_pure}
\end{equation}
which shares average number and \ac{QFI} with \( \rhoONN \).
However, as with \( \ket{\psiONs} \), this can be thought of as nothing more than a coherent method to probabilistically apply a N00N state.

\subsection{Implications of satisfying the convexity bound}

At this point we can be convinced that the metrological value of \( \rhoONs \) or \( \rhoONN \) is the (effective) probabilistic application of a N00N state.
Considering either the expansion of the tensor product in Eq.~\eqref{eq:psi_ons} or the (equal \ac{QFI}) effective state for phase-insensitive measurements in Eq.~\eqref{eq:rho_preselect}, the same appears to be true of \( \ket{\psiONs} \).
Here again we find ourselves dissatisfied with average number: we have associated the precision of our probe state directly with a single component of fixed number \( \onN \) which is detached from \( \avgN \) (with arbitrary \( \eta \)).

This is not to say that the probabilistic application of the N00N state is a valuable or even viable strategy compared to the deterministic application.
When vacuum is applied nothing is done and the measurement results are of no value---non-zero contributions to both average number and \ac{QFI} come solely from the N00N state component.
It is simply the interplay between the \ac{QFI} displaying a quadratic scaling in \( \onN \) while average number is only linear in \( \onN \) which---as noted previously in coherent~\cite{rivas_sub-heisenberg_2012} and incoherent~\cite{pezze_phase-sensitivity_2015,demkowicz-dobrzanski_quantum_2015} mixtures---gives rise to the arbitrarily large \ac{QFI} discussed in Sec.~\ref{sec:arbitrary}.

Indeed, if probabilistically preparing vacuum were a mechanism to genuinely increase precision while reducing sample exposure we could note that we can apply the vacuum as probe state and discard the (deterministic) measurement results without reducing the quality of the estimate.
However, if we are not using the measurement results and consider the vacuum state to have no effect on the sample~\footnote{It is possible that these vacuum runs could in practice be beneficial, offering time for the sample to recover and avoid photobleaching effects, or actively detrimental, exacerbating drift effects or extending exposure to detrimental environmental or background noise processes.
Ignoring any such noise processes and considering photons passing through the sample (rather than applying some model of absorption~\cite{perarnau-llobet_weakly_2021}) the vacuum state can be considered to have no effect on any sample.}
no practical distinction can be drawn between performing or simulating these vacuum applications.
This would open the the farcical situation that any Heisenberg-scaling scheme can be ``improved'' with respect to the average number by \emph{simulating} a classical mixture of the probe state and a vacuum probe.
Such improvement is patently false as the same useful applications of a Heisenberg-scaling probe state have simply been diluted with additiona zero-cost, zero-information probe states.
Instead, to avoid this loophole of guaranteed yet free arbitrary precision from any better than shot-noise probe(s), we should conclude that the cost of the classical mixture cannot be well-represented by the average number \( \avgN \).

\subsection{A more appropriate resource}\label{sec:newresource}

We have considered a number of different states: \( \ket{\psiNOON} \), \( \ket{\psiONs} \), \( \rhoONs \), \( \rhoONN \), and \( \ket{\psiONN} \), possessing various total number distributions, average number, and \ac{QFI}.
Yet in each instance the \ac{QFI} is that of the \( N \)-particle N00N state scaled by the probability of collapsing each state into that \( \onN \)-particle N00N state.
The states also share an identical \ac{SLD} operator as App.~\ref{app:mixed_qfi:noonlike} shows.

It is our view that these---and the associated mixed states generated by projecting in total number---all derive their metrological power solely from their possession of a N00N state component.
This makes the other---where \( \eta \ll 1 \), largely vacuum---components only relevant to the average number (for a given amplitude of the N00N state).
If one were to condition on events where a photon is detected (a necessary but not sufficient condition for such an event to be informative as to the phase shift) then, as Tab.~\ref{tab:conditioned_qfi} shows, the states with a vacuum-component appear no-better, if not worse.

\begin{table}[htb]
	\begin{ruledtabular}
	\begin{tabular}{ c C C C C }
		\( \rho \) & \avgN & \braket{0 | \rho | 0} & \qfi{\rho} & \frac{\qfi{\rho}}{1-\braket{0 | \rho | 0}} \\ \hline 
		\( \ket{\psiNOON} \) & N & 0 & N^2 & N^2 \\
		\( \ket{\psiONs} \), \( \rhoONs \) & \frac{2\eta^2\onN}{1+\eta^2} & \frac{1}{(1+\eta^2)^2} & \frac{2\eta^2\onN^2}{(1+\eta^2)^2} & \frac{\onN^2}{1+\frac{\eta^2}{2}} \\
		\( \ket{\psiONN} \), \( \rhoONN \) & \frac{2\eta^2\onN}{(1+\eta^2)^2} & \frac{1+\eta^4}{(1+\eta^2)^2} & \frac{2\eta^2\onN^2}{(1+\eta^2)^2} & \onN^2 \\
	\end{tabular}
	\end{ruledtabular}
	\caption{States, select properties of their total number distributions, and quantum Fisher information for the discussed pure states.}
	\label{tab:conditioned_qfi}
\end{table}

Rather than \( \eta \) being a parameter which can be tuned to increase the \ac{QFI}, it should be considered a parameter to reduce \( \avgN \) by obfuscating the N00N component solely responsible for the metrological power of all these states.
As such \( \onN \), not \( \avgN \), is a more representative resource for these states in terms of both the photons passing through the sample and their metrological performance.

\section{Probabilistic sensing}
\label{sec:prob}

The states of the previous example are particularly convenient to analyse, however, the fundamental conceit of doing nothing---by coherent or incoherent means---to artificially deflate average number can be applied more generally.
In both cases the average number of the resultant state becomes detached from the original state responsible for all phase-sensitivity giving a false impression of improved sensitivity per photon.

\subsection{Probabilistically sensing}
\label{sec:prob:mixed}

A two-mode state \( \ket{\phi_M} \) with average number \( M \) which is either used (with probability \( p \)) or replaced with vacuum (with probability \(1-p\))---without adapting the interferometer or measurement---as input to the interferometer is described by the state
\begin{equation}
    \rho = (1-p)\ketbra{0} + p \ketbra{\phi_M},
    \label{eq:prob_mix_probe}
\end{equation}
which---by definition of \( \ketbra{\phi_M} \)---contains on average \( p M \) photons.

The \ac{QFI} is (see App.~\ref{app:mixed_qfi:prob}) 
\begin{equation}
	\qfi{\rho} =
	4 p  \left( \braket{ \phi' | \phi' } - \abs{ \braket{ \phi' | \phi } }^2 \frac{1-p\abs{\braket{\phi | 0}}^2}{1-\abs{\braket{\phi|0}}^2} \right) 
	= p \qfi{\ket{\phi}} - \frac{p(1-p) \expect{\op{n}_1-\op{n}_2}^2 \abs{\braket{\phi | 0}}^2}{1-\abs{\braket{\phi | 0}}^2},
    \label{eq:probabilistic_qfi}
\end{equation}
which is upperbounded as \( \qfi{\rho} \leq p\qfi{\ket{\phi}} \), with the equality holding for any \( \ket{\phi} \) either orthogonal to the vacuum state (\( \braket{0 | \phi} = 0 \)), or symmetric in the two modes (\( \expect{\op{n}_1-\op{n}_2} = 0 \)).
Adapting the probability with which \( \ket{\phi_M} \) is used to fix \( p \) to be \( \expect{N} / M \)---%
for any \( \expect{N} \leq M \)---%
means the \ac{QFI} is at most
\begin{equation}
	\qfi{\rho} \leq \expect{N} \frac{\qfi{\ket{\phi_M}}}{M}.
    \label{eq:probabilistic_qfi_H}
\end{equation}

For states \( \ket{\phi^{\mathrm{S}}_M} \) which only reach a shot-noise scaling \( \qfi{\ket{\phi^{\mathrm{S}}_M}} \sim M \), Eq.~\eqref{eq:probabilistic_qfi_H} gives that \( \qfi{\rho^{\mathrm{S}}} \lesssim \expect{N} \)---classical states are constrained to only give a classical scaling when applied in such a probabilistic manner.
By comparison a Heisenberg scaling state possessing \( \qfi{\ket{\phi^{\mathrm{H}}_M}} \sim M^2 \), gives a \ac{QFI}
\(
    \qfi{\rho^{\mathrm{H}}_M} \lesssim \expect{N} M,
\)
which---at least when \( \braket{ \phi | 0 } \sim 0 \) or \( \expect{\op{n}_1-\op{n}_2} \sim 0 \)---can be taken to be arbitrarily large with suitably large \( M \), while---for any \( M > \expect{N} \)---\( p \) can be adjusted to achieve the fixed average number \( \expect{N} \).

Here it may also make sense to consider \( M \) as the more appropriate resource constraint for the states defined in Eq.~\eqref{eq:prob_mix_probe}.
With respect to \( M \), \( \qfi{\rho^{\mathrm{H}}_M} \) does not display any better-than-Heisenberg scaling and \( p \) acts again to reduce the average number without positively contributing to the metrological performance.
However, this relies on \( \ket{\phi_M} \) itself being well-represented by its average number; this fails (as it should) if \( \ket{\phi_M} = \ket{\psiONs} \).

\subsection{Coherent probabilistic applications}
\label{sec:prob:pure}

Now we turn to a coherent form of the probabilistic scheme where
\begin{equation}
	\ket{\psi} = \sqrt{1-p}\ket{0} + \sqrt{p} \ket{\phi_M}.
    \label{eq:prob_coh_probe}
\end{equation}
Limiting to \( \braket{ 0 \vert \phi_M } = 0 \) such that \( \ket{\psi} \) is normalised,
we have \( \op{n}_j \ket{\psi} = \sqrt{p} \op{n}_j \ket{\phi_M} \) and so the \ac{QFI} is
\begin{equation}
    \qfi{\ket{\psi}} = p \expect{(\op{n}_1-\op{n}_2)^2}_{\phi_M} - p^2 \expect{\op{n}_1-\op{n}_2}_{\phi_M}^2 .
    \label{eq:purified_qfi}
\end{equation}
This coherent probabilistic scheme can technically improve on the probabilistic scheme outlined in Sec.~\ref{sec:prob:mixed} by the relatively simple comparison of Eqs.~\eqref{eq:probabilistic_qfi} and~\eqref{eq:purified_qfi}, noting Eq.~\eqref{eq:purified_qfi} is greater \( p \qfi{\ket{\phi_M}} \) due to the additional factor $p$ reducing the \( \expect{\op{n}_1-\op{n}_2}_{\phi(N)}^2 \) term. 
Again, with a Heisenberg scaling \( \ket{\phi_M} \), \( \ket{\psi} \) can achieve a better-than-Heisenberg scaling with respect to \( \expect{N} \), but not \( M \).

\subsection{Average number}

In both the incoherent and coherent settings discussed, we can perform the same conditioning on events where a non-zero number of photons are detected (which is necessary but not sufficient for the event to be informative).
Considering \( \braket{0|\phi_M} = 0 \), photons are detected with probability \( p \) and precision can only increase when this happens.
Conditioning on such detection events---retaining \( p = \expect{N} / M \)---we find for \( \rho \)
\begin{equation}
	\frac{\qfi{\rho}}{p} = \qfi{\ket{\phi_M}} 
\end{equation}
and for \( \ket{\psi} \)
\begin{equation}
	\frac{\qfi{\ket{\psi}}}{p} = \qfi{\ket{\phi_M}} + (1-p)\expect{\op{n}_1-\op{n}_2}_{\phi_M}^2
\end{equation}
where the latter term can be upper-bounded by \( M^2 \) (as \( 1-p \leq 1 \) and \( \expect{\op{n}_1-\op{n}_2}^2 \leq \expect{\op{n}_1+\op{n}_2}^2 \)) which means for Heisenberg-scaling \( \ket{\phi_M} \) with \( \qfi{\phi_M} = M^2 \), \( \frac{\qfi{\ket{\psi}}}{p} \leq (2-p) \qfi{\phi_M} \).

\section{Bayesian analysis}
\label{sec:bayes}
In Sec.~\ref{sec:excess} we focused exclusively on the \ac{CRB} and \ac{QFI} as a tool to argue that certain variable-number states derive their metrological power from a common source, and that this fact can be obscured when the average number is employed as the resource. 
We now consider a Bayesian single-shot analysis to show that these observations transcend the local nature of the \ac{CRB}. 

\subsection{Metrological power in the presence of prior information}\label{sec:bayesmetro_power_theory}

Instead of a hierarchy of inequalities (e.g., Eq.~\eqref{eq:crbs}) which must be saturated sequentially in order to describe the attainable precision, the Bayesian framework enables a flexible formulation of optimisation problems whose solution, if available, readily provides the \emph{best} possible measurement and phase estimate~\footnote{%
While quantum metrology typically focuses on measurement strategies, quantum estimation \`{a} la Bayes can provide not only the optimal measurement and estimate for a given probe, as we do here, but also inform which initial probe is globally optimal \cite{macieszczak2014}.},
together with the associated uncertainty, for an appropriate criterion of performance~\cite{helstrom1976}.

Under the mean square error criterion~\cite{helstrom1976,demkowicz-dobrzanski_optimal_2011}, the optimal single-shot measurement is given by the eigenvectors of an operator $S$, solution to the equation $S \varrho + \varrho S = 2 \varrho'$, where $\varrho = \int_a^b d\varphi\, z(\varphi)\rho(\varphi)$, $\varrho' = \int_a^b d\varphi\,z(\varphi)\rho(\varphi)\varphi$, and $z(\varphi)$ is a prior probability with support bounded by $a$ and $b$~\cite{personick1971, rubio_quantum_2019}.
In this work, $\rho(\varphi) = \exp(-i \varphi K)\,\rho(0) \exp(i \varphi K)$, with $ K = (\hat{n}_1 - \hat{n}_2)/2$ and $\rho(0)$ the initial state fed to the interferometer.
Furthermore, the $j$-th eigenvalue of $S$ gives the optimal estimate for the phase shift $\varphi$ when the $j$-th projector is measured, and the optimal uncertainty is~\cite{personick1971,rubio_quantum_2019},
for a single shot~\footnote{Note that the form of Eq.~\eqref{eq:bayesian_opt_error} is still valid in multi-shot scenarios with collective measurements~\cite{rubio_quantum_2019}.}, 
\begin{equation}
    {\langle(\Delta \varphi)^2}\rangle_{\mathrm{opt}} = \int_a^b d\varphi\,z(\varphi)\varphi^2 - \mathrm{Tr}(\varrho S^2),
    \label{eq:bayesian_opt_error}
\end{equation}
where \( \langle(\Delta \varphi)^2\rangle \) denotes the prior-averaged mean square error~\cite{demkowicz-dobrzanski_quantum_2015,jarzyna_true_2015,li_frequentist_2018}.
The optimal phase estimates associated with Eq.~\eqref{eq:bayesian_opt_error} are generally biased, for, in Bayesian estimation, limiting the problem to unbiased estimators is not needed, nor necessarily beneficial. 
This is in contrast to earlier sections where \( (\Delta\varphi)^2 \) refers to the variance of an unbiased estimator, with unbiasedness being required there as biased estimators can display low variance despite, due to large bias, high error.

\footnotetext[6]{This follows from a linear expansion of \( \rho(\varphi) \) around the prior average \( \bar{\varphi} = \int_a^b d\varphi\,z(\varphi)\varphi \) as \( \rho(\varphi) \approx \rho(\bar{\varphi}) + \left. \frac{\partial \rho}{\partial \varphi} \right|_{\bar{\varphi}} (\varphi-\bar{\varphi}) \) which gives \( \varrho \approx \rho(\bar{\varphi}) \) and \( \varrho' \approx \bar{\varphi} \rho(\bar{\varphi}) + \sigma_0^2 \left. \frac{\partial \rho}{\partial \varphi} \right|_{\bar{\varphi}} \).
Under this expansion we find \( S = \bar{\varphi} + \sigma_0^2 \left. L_{\varphi} \right|_{\bar{\varphi}}  \) with \( L_{\varphi} \) being the \ac{SLD}, with which Eq.~\eqref{eq:bayesian_opt_error} becomes
\(
\langle(\Delta \varphi)^2\rangle_{\mathrm{opt}} \simeq \sigma_0^2 [1-\sigma_0^2\, \mathcal{J}(\rho)]
\)~\cite[App.~D]{rubio_quantum_2019}
}

\footnotetext[7]{Similar results have reported an exact equality for a Gaussian prior~\cite{macieszczak2014,demkowicz-dobrzanski_multi-parameter_2020}.}

Next, it is noted that the term $\mathrm{Tr}(\varrho S^2)$ is suggestive of the mixed-state definition for the \ac{QFI} (c.f.~App.~\ref{app:mixed_qfi}).
However, since the prior $z(\varphi)$ enters Eq.~\eqref{eq:bayesian_opt_error} non-trivially, a more careful examination is needed if we are to find a sensible quantifier of metrological power in Bayesian interferometry. 

If the prior is sufficiently narrow, Eq.~\eqref{eq:bayesian_opt_error} may reasonably be approximated as
\(
{\langle(\Delta \varphi)^2}\rangle_{\mathrm{opt}} \approx \sigma_0^2 [1-\sigma_0^2\, \mathcal{J}(\rho)]
\)~\cite{rubio_quantum_2019, Note6, Note7},
where $\mathcal{J}(\rho)$ is the \ac{QFI} employed so far and $\sigma_0^2 = \int_a^b d\varphi\,z(\varphi)\varphi^2 - [\int_a^b d\varphi\,z(\varphi)\varphi]^2$ is the initial uncertainty.
This motivates rewriting Eq.~\eqref{eq:bayesian_opt_error} in a similar form, but now without the aforementioned approximation, as 
\begin{equation}
{\langle(\Delta \varphi)^2}\rangle_{\mathrm{opt}} = \sigma_0^2 \left[1 - \sigma_0^2 \,\mathcal{P}(\rho; z) \right],
\end{equation}
where we have defined the quantity
\begin{equation}
    \mathcal{P}(\rho; z) \coloneqq \frac{\mathrm{Tr}(\varrho S^2) - \mathrm{Tr}(\varrho S)^2}{\sigma_0^4},
    \label{eq:genMP}
\end{equation}
and $\mathrm{Tr}(\varrho S) = \int_a^b d\varphi \, z(\varphi) \varphi$, with \( \rho \) and \( \varrho \) denoting the initial and the prior-averaged parameter-encoded states.
Note that Eq.~\eqref{eq:genMP} recovers the \ac{QFI} locally, i.e., $\mathcal{P}(\rho; z) \rightarrow \mathcal{J}(\rho)$ in the limit of a narrow prior. 
We now argue that $\mathcal{P}(\rho; z)$ is a valid and more general quantifier of metrological power. 

By construction, $0 \leq \mathcal{P}(\rho; z) \leq 1/\sigma_0^2$.
If $\mathcal{P}(\rho; z) = 0$, then ${\langle(\Delta \varphi)^2}\rangle_{\mathrm{opt}} = \sigma_0^2$, and so no information is gained by the application of the scheme. 
On the contrary, if $\mathcal{P}(\rho; z) = 1/\sigma_0^2$, then ${\langle(\Delta \varphi)^2}\rangle_{\mathrm{opt}} = 0$, which would imply that the relative phase is perfectly resolved.
Enhancing the precision is thus equivalent to reducing the prior error by making $\mathcal{P}(\rho; z)$ larger. 
Unlike the \ac{QFI}, \( \mathcal{P}(\rho; z) \) can, regardless of \( \rho \), only be unbounded when \( \sigma_0 = 0 \), where even an infinite amount of information would not improve the estimate since there is no uncertainty to be reduced. 

The quantifier $\mathcal{P}(\rho; z)$ thus possesses the properties that a good measure of metrological power should have, while generalising the \ac{QFI}. 
Specifically, in the limit where $\mathcal{P}(\rho; z) \rightarrow \mathcal{J}(\rho)$, Eq.~\eqref{eq:genMP} leads to the \emph{same} hierarchy of probes predicted by local estimation; otherwise, the Bayesian quantifier will generally reveal more accurate information about metrological enhancements. 

Note that, in general, one should use a criterion of performance that respects the periodicity of phase shifts~\cite{holevo2011, demkowicz-dobrzanski_quantum_2015}, and Eq.~\eqref{eq:bayesian_opt_error} does not.
Nevertheless, if the prior is effectively concentrated within an interval of length $W = (b-a) \lesssim 2$, then the square error giving rise to Eq.~\eqref{eq:bayesian_opt_error} approximates a 
\emph{sine} square error, which \emph{is} a valid cost function for periodic parameters~\cite{holevo2011, demkowicz-dobrzanski_quantum_2015, rubio_quantum_2019}. 
Admittedly, this restricts the quantifier $\mathcal{P}(\rho; z)$ to an intermediate regime between local estimation ($W \ll 1 $) and a fully global estimation framework ($W \leq 2\pi$).
Yet, such a regime has been shown to be sufficiently non-local (i.e., to allow for sufficiently wide priors) to reveal non-trivial effects beyond \ac{QFI} analyses in interferometry~\cite{rubio_quantum_2019}, and so it suffices for our purpose.

\subsection{Average number in Bayesian interferometry}\label{sec:avg_num_bayes}

We now demonstrate  that the decoupling of average number and the origin of the metrological power is not limited to local estimation.
To ensure equivalent prior knowledge in each case, we omit a direct analysis of \( \psiONs \) and \( \psiONN \),
and instead use their effective phase-insensitive measurement forms \( \rhoONs \) and \( \rhoONN \), which are also of independent interest viz. Sec.~\ref{sec:excess}.
App. \ref{app:bayes_calcs} provides the details of these calculations.

Consider a N00N state and prior $z(\varphi) = 1/W$, with support $\varphi \in [-W/2, W/2]$; the Bayesian quantifier (now omitting the prior dependency) from Eq.~\eqref{eq:genMP} reads
\begin{equation}
   \mathcal{P}(\ket{\psiNOON}) = \kappa(NW/2) N^2,
   \label{eq:bayes_noon_precision}
\end{equation}
where $\kappa(x)$ is function with range \( [0,1] \) and defined as $\kappa(x) = 9 \left(x\cos x - \sin x \right)^2/x^6$. The limit $\kappa \to 1$, which is realised when local prior information is available (i.e., \( W \to 0 \)), leads to \( \mathcal{P}(\ket{\psiNOON}) = N^2 \);
this not only is consistent with the \ac{QFI}, but also bounds \( \mathcal{P}(\ket{\psiNOON}) \) for any finite \( W \).
By fixing $NW=\zeta$, with finite $W$, we further recover the well-known result that N00N states achieve a Heisenberg scaling \begin{equation}
\mathcal{P}(\ket{\psiNOON}) = \kappa(\zeta/2)\,N^2 
\label{eq:noon_bayes_hl}
\end{equation}
subject to the prior being localised to an interval \( W = \zeta / N \).
Moreover, for \( \zeta \lesssim 1 \) we have \( \kappa \sim 1 \), implying that \( \mathcal{P}(\ket{\psiNOON}) \sim N^2\) when the phase is localised to an interval $W \sim 1/N$ \cite{berry_optimal_2012, hall_does_2012}. 

Now, compare this scheme with the variable-number state in Eq.~\eqref{eq:rho_preselect}, $\rho_{\psi_{0N}^{\otimes 2}}$, which, assuming the same flat prior, leads to
\begin{equation}
\mathcal{P}(\rho_{\psi_{0N}^{\otimes 2}}) = \frac{2\,\eta^2\,\mathcal{P}(\ket{\psiNOON})}{(1+\eta^2)^2}
= \frac{2\,\eta^2\,N^2\, \kappa(\zeta/2)}{(1+\eta^2)^2}.
\label{eq:gen_no_ref_beam_bayes}
\end{equation}
Recalling that, for this scheme, $\avgN = 2\,\eta^2 N/(1+\eta^2)$, Eq.~\eqref{eq:gen_no_ref_beam_bayes} can be written as
\begin{equation}
\mathcal{P}(\rhoONs) = \kappa(\zeta/2) \avgN \left( N - \frac{\avgN}{2} \right) \approx  \kappa(\zeta/2)\,\langle N \rangle N,
\label{eq:bayes_variable_num}
\end{equation}
where the approximation assumes that, for a fixed (and finite) $\avgN$, $N\gg \avgN$. 
In other words, $\rho_{\psi_{0N}^{\otimes 2}}$ cannot provide a precision scaling better than $\mathcal{P}(\rho_{\psi_{0N}^{\otimes 2}}) \sim \avgN N$, where $N$, and not $\avgN$, gauges the precision.
Moreover, the Heisenberg scaling is not violated with respect to $N$.

As with the \ac{QFI}, here too we are led to conclude that $N$ may be a more representative resource constraint.
This is further reinforced by the existence of a key relationship between $N$ and the prior width $W$---namely, $W\lesssim 1/N$---which is needed for $\mathcal{P}(\rho_{\psi_{0N}^{\otimes 2}}) \sim \avgN N$ to hold with a coefficient of order unity.
Given that the average number $\avgN$ plays no role in such relationship, using $\avgN$ as a proxy without examining the role of $N$ could lead to underestimating the prior information needed to exploit the metrological power of a state. 
For instance, we could be led to rely on \( W \sim 1 / \avgN \) which, should it be an insufficient amount of prior information, risks giving rise to exposure to a high-power probe state that does not extract information beyond the initial knowledge;
such a zero-gain in spite of a prior \( W \sim 1 / \avgN \) has been noted previously~\cite{rubio_bayesian_2020}.

Still, there is no doubt that $\rho_{\psi_{0N}^{\otimes 2}}$ has \emph{some} metrological value---\citet{luis_breaking_2017} also showed in the context of Bayesian estimation that the single-mode \( \ket{\psiONs} \) has some value---however, as in the \ac{QFI} discussion, its source may be linked, tentatively, to the N00N-state component.
First, note that the constraint $W\lesssim 1/N$ is a consequence of the form of the Bayesian precision for N00N states, regardless of the variable-number state to which the N00N-state component may belong to. 

Secondly, if we generalise the notion of resource introduced in Sec.~\ref{sec:newresource} from $\mathcal{J}(\rho)/p$ to $\mathcal{P}(\rho)/p$, where $p$ was the probability of detecting photons at the output ports, we find  
\begin{equation}
    \frac{\mathcal{P}(\rho_{\psi_{0N}^{\otimes 2}})}{p} = \frac{\mathcal{P}(\ket{\psiNOON})}{1+\frac{\eta^2}{2}} \leq \mathcal{P}(\ket{\psiNOON}),
    \label{eq:true_metro_power_gen_state}
\end{equation}
where we have used that, for $\rho_{\psi_{0N}^{\otimes 2}}$, $p = 1 - 1/(1+\eta^2)^2$.
Eq.~\eqref{eq:true_metro_power_gen_state} indicates that the sensitivity achieved by $\rho_{\psi_{0N}^{\otimes 2}}$ can never surpass that of a probabilistic application of its N00N-state component.
If we instead examine the mixed-N00N state $\rhoONN$ in Eq.~\eqref{eq:rho_probabilistc_on}, which removes the possibility of non-informative and potentially damaging events (i.e., those associated with $\ket{N N}$), then we have the exact identity
\begin{equation}
    \frac{\mathcal{P}(\rhoONN)}{p} = \mathcal{P}(\ket{\psiNOON}),
    \label{eq:true_metro_power_gen_state_exact}
\end{equation}
where now $p=2\eta^2/(1+\eta^2)^2$. 

Finally, in App. \ref{app:bayes_calcs} we show that---as with the \ac{SLD}---the quantum estimator $S$ is that for a N00N state even when $\rho_{\psi_{0N}^{\otimes 2}}$ or $\rhoONN$ are employed.
Given the fundamental role played by $S$ in Bayesian estimation---it contains the optimal phase estimates and their measurement scheme---this is perhaps the most convincing argument.

We conclude that the unsuitability of the average photon number to capture the true source of metrological advantage, as well as the real damage done to the sample, is not a byproduct of the local nature of the QFI, but a more general feature of quantum metrology.
More practically, Eqs.~\eqref{eq:noon_bayes_hl},~\eqref{eq:true_metro_power_gen_state} and~\eqref{eq:true_metro_power_gen_state_exact} extend the validity of the findings in Tab. \ref{tab:conditioned_qfi}, based on the \ac{QFI}, to a non-local regime with moderate prior knowledge (i.e., $W \lesssim 2$).

\section{Discussion}

\subsection{Understanding Heisenberg scaling with only an average number constraint}

We have seen that the close-to-vacuum states \( ( \ket{0} + \eta \ket{\phi} ) / \sqrt{1+\eta^2} \) appear to owe their metrological-advantage to being effective probabilistic application of Heisenberg-scaling states.
We would expect these observations to carry over to the likes of the Rivas-Luis states; Sec.~\ref{sec:prob:pure} does in principle cover them, but only through a state of the form \( \ket{\phi} = \ket{0,\xi} + \ket{\xi,0} + \alpha \ket{\xi,\xi} \) which we do not believe have garnered any direct analysis.
This already raises major questions as to whether the apparent Heisenberg-violating scalings are either: actually Heisenberg-violating, or quantum in origin.

An explicitly better-than-Heisenberg scaling requires \( N \) to be explicitly parameterised in terms of \( \eta \) and \( \avgN \), otherwise we simply have a highly favourable pre-factor to a \ac{QFI} linear in \( \avgN \) as Eq.~\eqref{eq:qfi_on_fixed_nbar}, or an unfavourable pre-factor to a \ac{QFI} quadratic in \( N \) as Eq.~\eqref{eq:qfi_on}.
Moreover, if we take \( N \) as our resource for these states, the parameterised form (e.g.~Eq.~\eqref{eq:qfi_on_fixed_nbar}) is bound by a Heisenberg-scaling in \( N \)---both in isolation and conditioning on photon-involving events (Tab.~\ref{tab:conditioned_qfi}).

Arbitrarily large precision gains are, in any case, explicitly forbidden when working with finite prior information (i.e., $\sigma_0$ is finite), since the Bayesian quantifier $\mathcal{P}(\rho; z)$ is upper-bounded by $1/\sigma_0^2$.
Even this finite improvement can be hard to achieve
with better-than-Heisenberg strategies requiring either a very large number of repetitions~\cite{tsang_ziv-zakai_2012, rubio2018} or a very large amount of prior knowledge~\cite{giovannetti_sub-heisenberg_2012, rubio_bayesian_2020}.
Moreover, there is an extensive body of evidence supporting that a Heisenberg-scaling in the total number used across all repetitions cannot be surpassed~\cite{braunstein_maxlikelihood1992, tsang_ziv-zakai_2012, berry_optimal_2012, pezze_phase-sensitivity_2015, gorecki2020}.

As to the quantum origin, coherence between different total number subspaces does not contribute to the precision and performance is no different to spending a fraction of trials using N00N states and a fraction doing nothing.
One can thus argue that the Heisenberg-scaling derives from the quintessentially quantum N00N state but (in the cases discussed herewanyhe apparent better-than-Heisenberg precision on top of this comes from the wholly classical ability to sometimes do nothing whether exactly or---due to coherence---only effectively.

\subsection{Noise}

Optical loss is known to be detrimental to precision in interferometry~\cite{demkowicz-dobrzanski_quantum_2009}.
Of particular relevance to this work is the inevitable reduction to a shot-noise scaling in sufficiently high loss~\cite{kolodynski_phase_2010, escher_general_2011, demkowicz-dobrzanski_elusive_2012}.
This gives rise to a bound on the \ac{QFI} of a variable number state \( \rho \) with average number \( \avgN \) of~\cite{demkowicz-dobrzanski_fundamental_2013}
\begin{equation}
   \qfi{\rho} \leq \frac{\avgN \gamma}{1-\gamma},
   \label{eq:loss_bound}
\end{equation}
for fractional loss rate \( \gamma \).
In low noise, Eq.~\eqref{eq:loss_bound} has a substantial pre-factor on \( \avgN \) which does not preclude Heisenberg-scaling (\( \gamma/(1-\gamma) \) can exceed \( \avgN \)), but for sufficiently high loss any scheme reduces to a shot-noise scaling.
In such a regime \( \avgN \) may suffice, though photodamage may remain a concern if the loss involves absorption or does not precede the sample.

In discussing the potential photodamage we limited our model to the number of photons passing through the probe. In practice, a more intricate model explicitly factoring in some absorption (damage) dynamic may be desirable both in quantifying a state's metrological performance and the potential damage it may cause by its application.
Whether a given level of exposure, or absorption or damage is considered can give rise to different precisions~\cite{wolfgramm_entanglement-enhanced_2013,perarnau-llobet_weakly_2021}.

\subsection{Beyond an average number constraint}

Although Heisenberg-scaling is ordinarily identified as a quadratic scaling in~\( \avgN \)~\cite{ou_fundamental_1997,luis_nonlinear_2004,lang_optimal_2014,sahota_quantum_2015}, previous works have identified a linear scaling in \( \avgSqN \) (quadratic in \( \sqrt{\avgSqN} \)) as the correct fundamental limit when variable number states are used in interferometry~\cite{hofmann_all_2009,hyllus_entanglement_2010,pezze_phase-sensitivity_2015,gessner_sensitivity_2018}.
For \( \ket{\psiONs} \), which has \( \avgSqN = 2 \eta^2 (1+2\eta^2) N^2 / (1+\eta^2)^2 \), the QFI indeed scales linearly in \( \avgSqN \) with \( \qfi{\ket{\psiONs}} / \avgSqN = 1 / (1+2\eta^2) \).
The coefficient \( \qfi{\ket{\psiONs}} / \avgSqN \) is of order unity in the Heisenberg-violating \( \eta \ll 1 \) regime and vanishing---as the \ac{QFI} itself vanishes---in the \( \eta \gg 1 \) regime where the input approximates phase-insensitive Fock states incident directly on the phase shifts.
Similarly, one may use Eqs.~(\ref{eq:bayes_noon_precision}--\ref{eq:gen_no_ref_beam_bayes}) to show that, in the Bayesian framework, $\mathcal{P}(\rho_{\psiONs})/\avgSqN \sim 1/(1+2\eta^2)$.

One could consider using \( \sqrt{\avgSqN} \) rather than or in addition to \( \avgN \) as a constraint; this gives a much larger cost for \( \ket{\psiONs} \) that better matches the average photon number when conditioned on events where a non-zero number of photons is detected.
For fixed number and coherent states \( \sqrt{\avgSqN} = \avgN \), while \( \sqrt{\avgSqN} \approx \avgN \) for squeezed vacuum states~\cite{loudon_squeezed_1987,lang_optimal_2014}.
This allows the existing shot-noise and Heisenberg scalings to carry over for those states under a \( \avgSqN \) constraint, while marking out the vacuum-Fock or Rivas-Luis states as much more expensive than average number alone suggests.

Working to an additional constraint of \( \avgSqN \) may make more sense to understand the potential exposure to a sample, particularly in the Heisenberg-violating regimes we explored earlier, however it is not sufficient to alleviate worst-case concerns.
Consider a superposition of primarily squeezed vacuum with a vanishing component of a much higher energy Fock state, in some sense an inverse situation to \( \ket{\psiONs} \); we would expect to find an apparently reasonable sensitivity: quadratic in \( \avgN \) and \( \sqrt{ \avgSqN } \) due to the squeezed vacuum contribution dominating, but still possessing a non-zero probability of a potentially damaging event.
While this situation would be better for bounding \( \avgN^2 \) and \( \avgSqN \), we can still have a rare, damaging, yet uninformative event---whether in terms of the number of photons passing through the sample or explicitly modelling absorption.
Simply constraining the maximum number---essentially truncating the Fock space by total photon number---goes a long way to alleviate this, but such a truncation may be stricter than need be, forbidding for example the conventional Gaussian states.

Instead, there is still some cause to desire a more intricately weighted measure of information per cost---such as an extension of the quantities \( \qfi{\ket{\psi}}/(1-\abs{\braket{0|\psi}}^2) \) and \( \mathcal{P}({\ket{\psi};z})/(1-\abs{\braket{0|\psi}}^2) \) employed in Sec.~\ref{sec:newresource} and Sec.~\ref{sec:avg_num_bayes}, respectively---capable of ensuring good sensing value per-photon in every aspect of the probe state.
More general measures of fluctuations in the total number distribution may be sufficient for this purpose.
Even simple heuristics may be of use: the total number distribution of Gaussian or fixed-number states are singly-peaked (around \( \avgN \)) while the vacuum-Fock and Rivas-Luis states are multiply-peaked (around $0$, $x$, and $2x$, with $x$ being the $N$ of the Fock state and the average number of the squeezed vacuum state).

\section{Conclusions}

We have argued that average number can fail to capture properties of the full total number distribution of probe states relevant to metrological power, photodamage, and necessary prior information.
While average number has been known not to properly constrain the theoretical precision of a probe state~\cite{hofmann_all_2009,pezze_phase-sensitivity_2015}, we argue this should be attributed to an over-reliance on average number to capture a potentially complex, multi-faceted number distribution rather than the existence of certain anomalous states.

Finally, we emphasise that while the examples considered here are informative as to the role average number should or should not be allowed to play in evaluating and comparing interferometry schemes, this does not negate the practical inaccessibility of supposed better-than-Heisenberg scaling states~\cite{tsang_ziv-zakai_2012,giovannetti_sub-heisenberg_2012,hall_universality_2012,berry_optimal_2012,pezze_sub-heisenberg_2013,jarzyna_true_2015}, nor can it circumvent the value of states and measurements which are practical to prepare~\cite{thomas-peter_integrated_2011} and noise-resilient~\cite{datta_quantum_2011}.

\begin{acknowledgments}
We are grateful to Francesco Albarelli for valuable discussions.
We thank Erik Gauger and Alfredo Luis for comments.
DB acknowledges support from UK EPSRC Grant EP/R030413/1.
JR acknowledges support from UK EPSRC Grant No.~EP/T002875/1.
\end{acknowledgments}

\appendix

\section{Precision calculations}

\subsection{Mixed state QCRB}
\label{app:mixed_qfi}

For a mixed state \( \rho \) the \acs*{SLD} \ac{QFI} is not given directly by Eq.~\eqref{eq:phase_qfi} but can be found as~\cite{braunstein_statistical_1994,paris_quantum_2009}
\begin{equation}
    \qfi[\varphi]{\rho} = \Tr (\rho L_{\varphi}^2) ,
    \label{eq:qfi_sld}
\end{equation}
where \( L_{\varphi} \) is the the \ac{SLD} operator defined by
\begin{equation}
    \frac{\partial \rho}{\partial \varphi} = \frac{ L_{\varphi} \rho + \rho L_{\varphi} }{2}.
    \label{eq:sld}
\end{equation}

\subsubsection{Mixed state relatives of the N00N and vacuum-Fock states}
\label{app:mixed_qfi:noonlike}

In Sec.~\ref{sec:excess} we have parameter-dependent mixed states of the form
\begin{equation}
    \rho(\varphi) =\, \alpha \ketbra{00} + \beta \ketbra{\psiNOON(\varphi)}
    + (1-\alpha-\beta)\ketbra{NN},
    \label{eq:master-state}
\end{equation}
whose derivative is
\begin{equation}
\partial_{\varphi} \rho =
     i\frac{\beta N}{2} e^{iN\varphi} \ket{N0}\bra{0N}
    -i\frac{\beta N}{2} e^{-iN\varphi} \ket{0N}\bra{N0},
\end{equation}
and so the \ac{SLD} is
\begin{equation}
    L_{\varphi} =
    i N ( e^{iN\varphi} \ket{N0}\bra{0N} - e^{-iN\varphi} \ket{0N}\bra{N0} ),
    \label{eq:app:sld}
\end{equation}
which is also the \ac{SLD} operator for any pure state where \( \ket{\psiNOON} \) is the only phase-sensitive component as \( L_{\varphi} \ket{\psiNOON(\varphi)} = \partial_{\varphi} \ket{\psiNOON(\varphi)} \).
The \ac{QFI} for Eq.~\eqref{eq:master-state} is then
\begin{equation}
    \qfi{\rho} = \beta N^2,
\end{equation}
which generalises the specific mixed state \acp{QFI} given in Sec.~\ref{sec:excess}.

\subsubsection{Probabilistic sensing}
\label{app:mixed_qfi:prob}

In Sec.~\ref{sec:prob} we consider the state Eq.~\eqref{eq:prob_mix_probe}, whose \ac{QFI} is most conveniently derived using a non-orthogonal basis~\cite{genoni_non-orthogonal_2019}.
With the non-orthognal basis \( \{ \ket{0}, \ket{\phi_M}, \ket{\phi'_M} \} \), where \( \ket{\phi'_M} = \partial_{\varphi} \ket{\phi_M} \), the \ac{SLD} solution is
\begin{equation}
\begin{aligned}
    L_{\varphi} =
    \frac{2}{p(1-\abs{\braket{0|\phi_M}}^2)} 
   \bigg[ &
    p\braket{\phi'_M|\phi_M} \braket{0|\phi_M} \ket{0}\bra{\phi_M}
    + p\braket{\phi_M|\phi'_M} \braket{\phi_M|0} \ket{\phi_M}\bra{0}
    \\
    &- \braket{0|\phi_M} \ket{0}\bra{\phi'_M}
    - \braket{\phi_M|0} \ket{\phi'_M}\bra{0}
    + \ket{\phi'_M}\bra{\phi_M}
    + \ket{\phi_M}\bra{\phi'_M}
   \bigg],
\end{aligned}
\end{equation}
recalling \( \braket{\phi_M|\phi'_M} + \braket{\phi'_M|\phi_M} = 0 \)~\cite{paris_quantum_2009} and \( \braket{0|\phi'_M} = 0 \) as \( \braket{0|\phi_M(\varphi)} = \braket{0|e^{i\varphi(\op{n}_1-\op{n}_2)}|\phi_M(0)} = \braket{0|\phi_M(0)} \).
The \ac{QFI} is then
\begin{equation}
    \qfi{\rho} = 4p(\braket{\phi'_M|\phi'_M} + \braket{\phi_M|\phi'_M}^2) + 4p(1-p) \frac{\abs{\braket{\phi_M|0}}^2 \braket{\phi'_M|\phi_M}^2}{1-\abs{\braket{\phi_M|0}}^2},
\end{equation}
which is upperbounded by \( 4p(\braket{\phi'_M|\phi'_M} + \braket{\phi_M|\phi'_M}^2) \) as \( \braket{\phi_M|\phi'_M} + \braket{\phi'_M|\phi_M} = \braket{\phi_M|\phi'_M} + \braket{\phi_M|\phi'_M}^* = 0 \) entails \( \braket{\phi_M|\phi'_M}^2 \leq 0 \).
Finally, \( \ket{\phi'_M} = i\frac{\op{n}_1-\op{n}_2}{2}\ket{\phi_M} \) gives Eq.~\eqref{eq:probabilistic_qfi}.

\subsection{Bayesian quantifier}\label{app:bayes_calcs} 

Sec. \ref{sec:bayesmetro_power_theory} introduces a measure of metrological power, $\mathcal{P}(\rho; z) = [\mathrm{Tr}(\varrho S^2) - \mathrm{Tr}(\varrho S)^2]/\sigma_0^4$, capable of accommodating a finite and moderate amount of prior information. 
Given the flat prior in the main text, i.e., $z(\varphi) = 1/ W$, when $\varphi \in [-W/2, W/2]$, such quantifier becomes
\begin{equation}
    \mathcal{P}(\rho) = \frac{144\,\mathrm{Tr}(\varrho S^2) }{W^4},
    \label{eq:metro_power_app}
\end{equation}
where we have used that $\sigma_0^2 = W^2/12$ and
\begin{equation}
    \mathrm{Tr}(\varrho S) =\int_a^b d\varphi\, z(\varphi)\, \varphi \propto \int_{-W/2}^{W/2} d\varphi\,\varphi = 0.
\end{equation}
The optimal quantum estimator $S$ is given by $S \varrho + \varrho S = 2 \varrho'$, where, in this case, 
\begin{equation}
\begin{aligned}
    \varrho &= \frac{1}{W}\int^{W/2}_{-W/2} d\varphi\,\rho (\varphi), &
    \varrho' &= \frac{1}{W}\int^{W/2}_{-W/2} d\varphi\,\rho (\varphi)\,\varphi.
\end{aligned}
\label{eq:bayes_aux_ops}
\end{equation}
As the SLD in Eq.~\eqref{eq:sld}, $S$ is determined by a Lyapunov equation. 
However, the similarity ends here: unlike in the frequentist approach, in the Bayesian framework we integrate the phase-dependent state with respect to the prior, making the final solution parameter-independent, and the derivative $\partial_\varphi\rho$ no longer plays a role (although it \emph{does} reemerge in the limit of a narrow prior, as one would expect from a local approach \cite{rubio_quantum_2019,Note6}).

Consider now the family of parameter-dependent states in Eq.~\eqref{eq:master-state}. The operators in Eq.~\eqref{eq:bayes_aux_ops} read
\begin{equation}
    \varrho = \, \alpha \ketbra{00} + (1-\alpha-\beta)\ketbra{NN}
    + \frac{\beta\,s_{NW}}{N W}(\ket{0N}\bra{N0}+\ket{N0}\bra{0N})  
    +\frac{\beta}{2} (\ketbra{0N}+\ketbra{N0})
    \label{eq:bayes_zeroth_moment}
\end{equation}
and
\begin{equation}
        \varrho' = i\,\frac{\beta\,(N W\,c_{NW} - 2\,s_{NW})}{2 N^2 W}\,(\ket{0N}\bra{N0}-\ket{N0}\bra{0N}),
\end{equation}
where $s_{NW} = \sin(N W/2)$ and $c_{NW} = \cos(N W/2)$. The solution to $S \varrho + \varrho S = 2 \varrho'$ is then
\begin{equation}
    S = i\, \frac{N W\,c_{NW} - 2\,s_{NW}}{N^2 W} \,(\ket{0N}\bra{N0}-\ket{N0}\bra{0N}).
    \label{eq:op_quantum_est}
\end{equation}
Introducing Eqs.~\eqref{eq:bayes_zeroth_moment} and \eqref{eq:op_quantum_est} in Eq.~\eqref{eq:metro_power_app}, we finally arrive at 
\begin{equation}
    \mathcal{P}(\rho) = \frac{144 \beta \left(N W c_{NW} - 2\,s_{NW} \right)^2}{N^4 W^6} = \kappa(N W / 2) \beta N^2,
    \label{eq:metro_power_app_solution}
\end{equation}
where we have defined the function $\kappa(x) = 9 \left(x\cos x - \sin x \right)^2/x^6$. The results discussed in Sec.~\ref{sec:avg_num_bayes} can be then recovered from Eq.~\eqref{eq:metro_power_app_solution} as follows: $\beta = 1$ for Eq.~\eqref{eq:bayes_noon_precision} (N00N state), and $\beta = 2\eta^2/(1+\eta^2)^2$ for both Eq.~\eqref{eq:gen_no_ref_beam_bayes} (state $\rhoONs$) and  $\mathcal{P}(\rhoONN)$ in Eq.~\eqref{eq:true_metro_power_gen_state_exact}. 

We further note that the optimal quantum estimator $S$ in Eq.~\eqref{eq:op_quantum_est} is independent of $\alpha$ and $\beta$, and so identical for all the states represented by Eq.~\eqref{eq:master-state}, which include the N00N state. This leads to the third argument given in Sec.~\ref{sec:avg_num_bayes} to support the idea that the metrological advantage of states such as $\rhoONs$ and $\rhoONN$, both included in Eq.~\eqref{eq:master-state}, originates in their N00N-state components. 

\bibliography{avg}

\end{document}